\documentclass[journal]{IEEEtran}
\usepackage{graphicx,amsmath,array,subfigure,url,booktabs,xcolor,multirow,color,balance,amsfonts,algorithmic,algorithm,textcomp,stfloats,verbatim,newclude,accents,amssymb,amsthm,cite,placeins,bm}
\begin{document}

\title{Polarforming Design for Movable Antenna Systems}

\author{
	Zijian Zhou, \IEEEmembership{Member,~IEEE},
	Jingze Ding, \IEEEmembership{Graduate Student Member,~IEEE}, and
  Rui Zhang, \IEEEmembership{Fellow,~IEEE}
	% <-this % stops a space
%	\thanks{\emph{(Corresponding authors: Jingze Ding; Rui Zhang.)}}
	\thanks{Zijian Zhou is with the School of Science and Engineering, The Chinese University of Hong Kong, Shenzhen 518172, China (e-mail: zijianzhou@link.cuhk.edu.cn).}
	\thanks{Jingze Ding is with the School of Electronics, Peking University, Beijing 100871, China (e-mail: djz@stu.pku.edu.cn).}
	\thanks{Rui Zhang is with the School of Science and Engineering, Shenzhen Research Institute of Big Data, The Chinese University of Hong Kong, Shenzhen 518172, China (e-mail: rzhang@cuhk.edu.cn). He is also with the Department of Electrical and Computer Engineering, National University of Singapore, Singapore 117583 (e-mail: elezhang@nus.edu.sg).}}
\maketitle

\begin{abstract}
Polarforming has emerged as a promising technique to enable the antenna to shape its polarization into a desired state for aligning with that of the received electromagnetic (EM) wave or reconfiguring that of the transmitted EM wave.  In this letter, we investigate polarforming design for the movable antenna (MA)-enabled communication system.  Specifically, we consider a single-input single-output (SISO) system with reconfigurable antenna positions and polarizations to leverage both spatial and polarization degrees of freedom (DoFs).  First, we present a polarized channel model and characterize the channel response as a function of antenna positions and polarforming phase shifts.  To maximize the achievable rate of the proposed system, we then develop a successive convex approximation (SCA)-based optimization algorithm by iteratively optimizing the antenna positions and phase shifts at both the transmitter and receiver.  Furthermore, simulation results demonstrate the performance gains of the proposed system over conventional systems in mitigating channel depolarization and adapting to channel fading.
\end{abstract}

\section{Introduction} \label{sec1}
\IEEEPARstart{M}{ovable} antenna (MA) technology is a promising solution for fully unlocking spatial degrees of freedom (DoFs) in wireless communication systems \cite{ZMZ24Jun, SJZ25Jan}.  Unlike traditional fixed-position antennas (FPAs), MAs can be dynamically repositioned/rotated within an up to three-dimensional space using drivers such as stepper motors.  By leveraging multi-path superposition in wireless channels, MA-aided systems offer additional spatial DoFs compared to conventional multiple-input multiple-output (MIMO) systems with FPAs.  While the advantages of MAs have been extensively explored, current MA systems are approaching their theoretical limits in exploiting spatial DoFs.  To meet the ever-growing demand for higher data rates, it is imperative to further explore new DoFs from new aspects such as polarization.  However, existing research works on MA (e.g., \cite{MWN24Jul, DZJ25Mar, LYX25Apr, DZZ25Jan, HWL25Mar}) mainly consider unpolarized channel models and thus overlook the untapped potential of polarization DoFs.

To capture polarization DoFs, we adopt polarforming to enhance the performance of MA systems in this letter.  Polarforming \cite{ZDW24Sep} enables dynamic polarization adjustment of antennas to reshape electromagnetic (EM) waves into a desired polarization state, thus effectively mitigating polarization loss caused by antenna misalignment and channel depolarization \cite{DZS25May}.  Different from existing polarforming techniques that leverage antenna translation/rotation \cite{ZSE24Nov, SZZ25May} or dual-polarized antennas (DPAs) \cite{ZDW24Sep, OB23Feb}, we employ the low-cost phase shifter (PS)-based polarization-reconfigurable antennas (PRAs) \cite{ZDW24Sep}, where each antenna integrates a single PS for enabling polarforming.  By manipulating phase shifts, this cost-effective design can achieve linear, circular, or general elliptical polarization to combat channel fluctuations and depolarization effects.

In this letter, we investigate polarforming design for the MA-enabled communication system, where each antenna employs reconfigurable position and polarization to jointly exploit both spatial and polarization DoFs.  Specifically, we consider a single-input single-output (SISO) system that enables flexible antenna movement and polarization control.  For this system, we first present a polarized channel model by characterizing the channel response as a function of antenna positions and polarforming phase shifts.  Next, we develop a successive convex approximation (SCA)-based algorithm to jointly optimize the antenna positions and phase shifts at both the transmitter and receiver for maximizing the system achievable rate.  Finally, simulation results are provided to demonstrate the performance gains and benefits of the proposed system over conventional systems with fixed antenna positions and/or polarizations.

\textit{Notations}: $a$, $\mathbf{a}$, $\mathbf{A}$, and $\mathcal{A}$ denote a scalar, a vector, a matrix, and a set, respectively.  $(\cdot)^T$ and $(\cdot)^H$ denote the transpose and conjugate transpose, respectively. $\mathbb{R}^{p\times q}$ and $\mathbb{C}^{p\times q}$ represent the spaces of $p\times q$ real-valued and complex-valued matrices, respectively.  $\otimes$ denotes the Kronecker product, and $\odot$ denotes the Hadamard product.  $\angle{x}$ and $|x|$ denote the phase and absolute value of a complex number $x$, respectively.  For a matrix ${\bf A}$, $[{\bf A}]_{pq}$ denotes the entry in the $p$-th row and $q$-th column.  ${\bf B} = {\rm blkdiag}\{{\bf b}_1,{\bf b}_2,\cdots,{\bf b}_L\}$ denotes a block diagonal matrix composed of the blocks ${\bf b}_1,{\bf b}_2,\cdots,{\bf b}_L$ along the diagonal.  $\mathbf{I}_K$ denotes an identity matrix of order $K$.  $\in$ stands for ``belong(s) to'', and $\succeq$ denotes generalized inequality.

\section{System Model and Problem Formulation} \label{sec2}
\begin{figure}[!t]
	\centering
	\includegraphics[width=\linewidth]{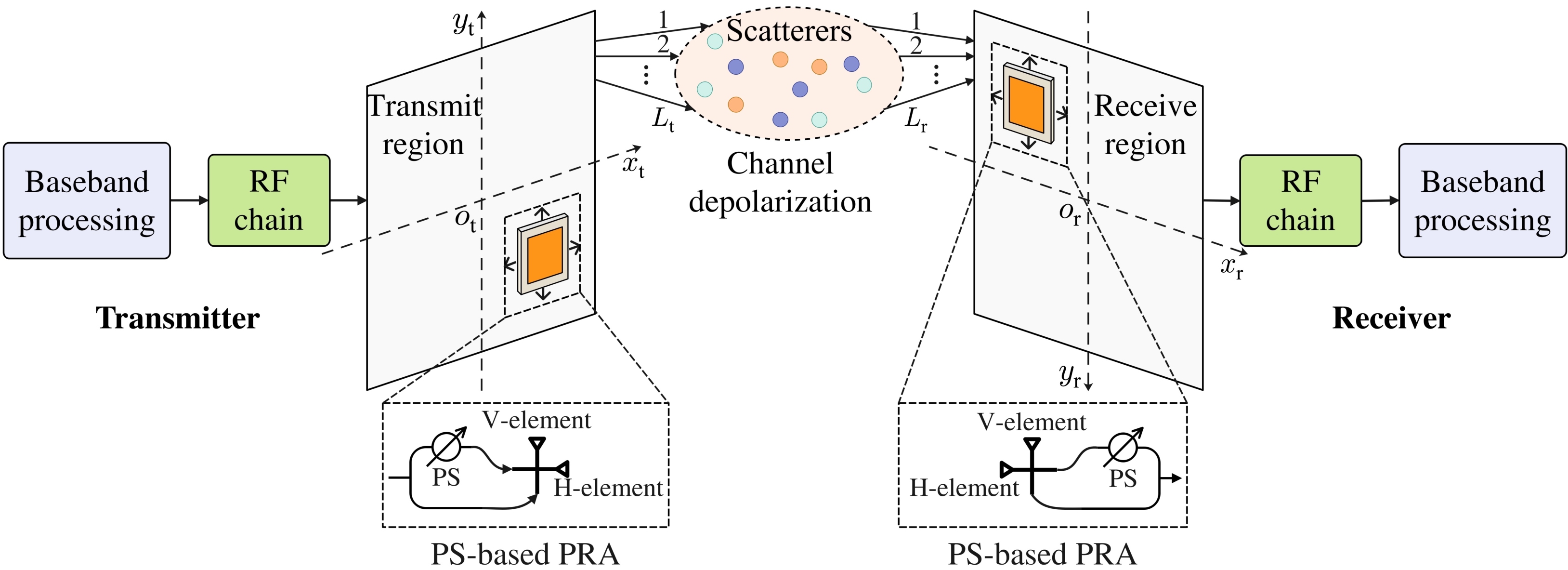}
	\caption{A schematic of polarforming design for the MA-enabled communication system.}
	\label{fig_system_model}
\end{figure}

As depicted in Fig.\;\ref{fig_system_model}, we consider a SISO system with reconfigurable antenna position and polarization.  The transmit/receive antenna consists of two orthogonal antenna elements, i.e., V-element for vertical polarization and H-element for horizontal polarization, and is also equipped with a stepper motor for flexible antenna movement.  A PS is integrated into each antenna front-end to control the phase difference between the two antenna elements for dynamic polarization control.  Let ${\bf t} = [x_{\rm t}, y_{\rm t}]^T \in {\cal C}_{\rm t}$ and ${\bf r} = [x_{\rm r}, y_{\rm r}]^T \in {\cal C}_{\rm t}$ represent the antenna position vectors (APVs) at the transmitter and receiver, respectively.  The corresponding phase shifts are denoted by $\theta \in [0,2\pi]$ for the transmitter and $\phi \in [0,2\pi]$ for the receiver.  Without loss of generality, we assume that the moving regions are squares, i.e., ${\cal C}_{\rm t} = [-\frac{A_{\rm t}}{2}, \frac{A_{\rm t}}{2}]^2$ and ${\cal C}_{\rm r} = [-\frac{A_{\rm r}}{2}, \frac{A_{\rm r}}{2}]^2$, where $A_{\rm t}$ and $A_{\rm r}$ are the sizes of the moving regions.

\begin{figure}[!t]
	\centering
	\includegraphics[width=0.65\linewidth]{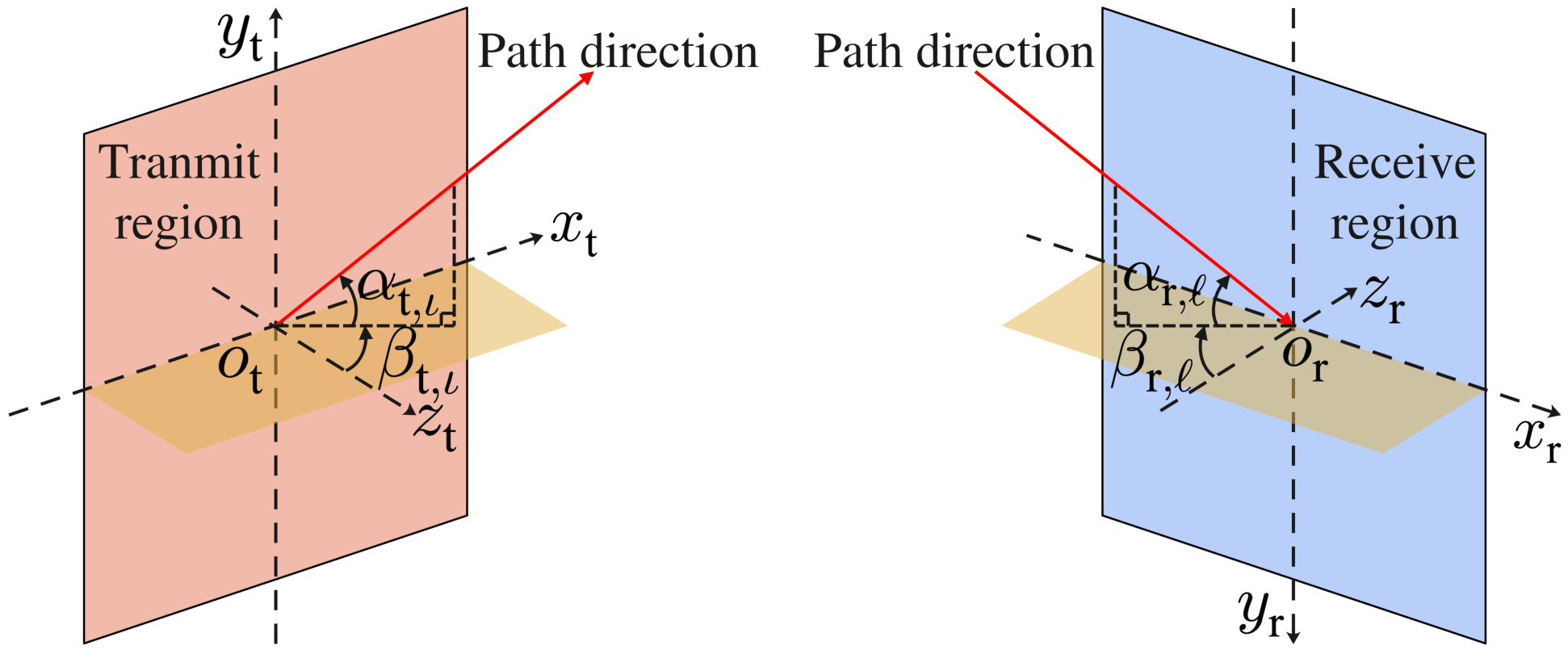}
	\caption{Illustration of the coordinates and spatial angles for the transmit and receive regions.}
	\label{fig_channel_model}
\end{figure}

For the considered system, we assume narrow-band quasi-static channels where the channel response depends on both the APVs and phase shifts, i.e., $h({\bf t},{\bf r},\theta,\phi)$.  Consequently, the received signal can be expressed as
\begin{equation}
	y({\bf t},{\bf r},\theta,\phi) = h({\bf t},{\bf r},\theta,\phi) P_{\rm t} s + z,
\end{equation}
where $s\in\mathbb{C}$ is the transmitted signal with zero mean and normalized power of one, $P_{\rm t}$ is the transmit power, and $z$ is the additive white Gaussian noise at the receiver with an average noise power of $\sigma^2$.  It is assumed that the noise from antenna elements and PSs is negligible compared to radio frequency (RF) chain noise, as the latter dominates in practice due to active components in the RF chains.

As illustrated in Fig.\;\ref{fig_channel_model}, let $\alpha_{{\rm t},\iota},\beta_{{\rm t},\iota}\in [-\frac{\pi}{2},\frac{\pi}{2}]$ denote the elevation and azimuth angles of departure for the $\iota$-th transmit path, and $\alpha_{{\rm r},\ell},\beta_{{\rm r},\ell}\in [-\frac{\pi}{2},\frac{\pi}{2}]$ represent the corresponding angles of arrival for the $\ell$-th receive path, where $1\le \iota \le L_{\rm t}$ and $1\le \ell \le L_{\rm r}$, with $L_{\rm t}$ and $L_{\rm r}$ being the numbers of transmit and receive paths, respectively.  Thus, we have $\rho_{{\rm t},\iota}({\bf t}) = x_{\rm t} \cos\alpha_{{\rm t},\iota}\sin\beta_{{\rm t},\iota} + y_{\rm t} \sin\alpha_{{\rm t},\iota}$ and $\rho_{{\rm r},\ell}({\bf r}) = x_{\rm r} \cos\alpha_{{\rm r},\ell}\sin\beta_{{\rm r},\ell} + y_{\rm r} \sin\alpha_{{\rm r},\ell}$.  According to \cite{ZMZ24Jun}, the field-response vectors (FRVs) at the transmitter and receiver are respectively defined as
\begin{align}
	{\bf u}({\bf t}) &= \left[ e^{j\frac{2\pi}{\lambda} \rho_{{\rm t},1}({\bf t})}, e^{j\frac{2\pi}{\lambda} \rho_{{\rm t},2}({\bf t})}, \cdots, e^{j\frac{2\pi}{\lambda} \rho_{{\rm t},L_{\rm t}}({\bf t})} \right]^T \in \mathbb{C}^{L_{\rm t}}, \\
	{\bf v}({\bf r}) &= \left[ e^{j\frac{2\pi}{\lambda} \rho_{{\rm r},1}({\bf r})}, e^{j\frac{2\pi}{\lambda} \rho_{{\rm r},2}({\bf r})}, \cdots, e^{j\frac{2\pi}{\lambda} \rho_{{\rm r},L_{\rm r}}({\bf r})} \right]^T \in \mathbb{C}^{L_{\rm r}},
\end{align}
where $\lambda$ represents the wavelength.  We then define the transmit and receive polarforming vectors (PFVs) using their respective phase shifts to characterize the antenna polarization states as \cite{ZDW24Sep}
\begin{equation}
	{\bf p}(\theta) = \frac{1}{\sqrt{2}}\begin{bmatrix} 1\\e^{j\theta} \end{bmatrix}, ~{\bf q}(\phi) = \begin{bmatrix} 1\\e^{j\phi} \end{bmatrix}.
\end{equation}
Since the effective channel response is a superposition of all paths, it can be expressed as
\begin{align} \label{sum_h}
	h({\bf t}, {\bf r}, \theta, \phi) &= \sum_{\iota = 1}^{L_{\rm t}} \sum_{\ell=1}^{L_{\rm r}} e^{-j\frac{2\pi}{\lambda}\rho_{{\rm r}, \ell}({\bf r})} {\bf q}(\phi)^H {\bf \Lambda}_{\ell\iota} {\bf p}(\theta) e^{j\frac{2\pi}{\lambda}\rho_{{\rm t}, \iota}({\bf t})} \nonumber\\
	&= {\bf g}({\bf r},\phi)^H {\bf \Lambda} {\bf f}({\bf t},\theta),
\end{align}
where ${\bf f}({\bf t},\theta) = {\bf u}({\bf t}) \otimes {\bf p}(\theta)$ and ${\bf g}({\bf r},\phi) = {\bf v}({\bf r}) \otimes {\bf q}(\phi)$.  The matrix ${\bf \Lambda}$ represents the path polarization response matrix (PPRM), defined as
\begin{equation} \label{pprm}
	{\bf \Lambda} = \begin{bmatrix}
	{\bf \Lambda}_{11} & \cdots & {\bf \Lambda}_{1L_{\rm t}} \\
	\vdots & \ddots & \vdots \\
	{\bf \Lambda}_{L_{\rm r}1} & \cdots & {\bf \Lambda}_{L_{\rm r}L_{\rm t}}
	\end{bmatrix} \in \mathbb{C}^{2L_{\rm r} \times 2L_{\rm t}}.
\end{equation}
Each submatrix ${\bf \Lambda}_{\ell\iota}\in \mathbb{C}^{2\times 2}$ represents the two-by-two polarized channel matrix between the $\iota$-th transmit and $\ell$-th receive path.

To evaluate the theoretical performance limit of the considered system, we assume perfect channel state information of the PPRM given in \eqref{pprm} is available at both the transmitter and receiver.  The achievable rate of the considered system in bits-per-second-per-Hertz (bps/Hz) is thus given by
\begin{equation} \label{rate}
  R({\bf t}, {\bf r}, \theta, \phi) = \log_2\left(1 + \frac{\left| h({\bf t}, {\bf r}, \theta, \phi) \right|^2 P_{\rm t}}{\sigma^2}\right).
\end{equation}
Our goal is to maximize the achievable rate in \eqref{rate} through joint optimization of the transmitter and receiver antenna positioning and polarforming variables $\{ {\bf t},{\bf r},\theta,\phi\}$.  Observing that $R({\bf t}, {\bf r}, \theta, \phi)$ in \eqref{rate} is a strictly increasing function over the channel power gain $\left| h({\bf t}, {\bf r}, \theta, \phi) \right|^2$, we can equivalently formulate the optimization problem as
\begin{subequations}
	\label{opt_problem}
	\begin{align}
		\max_{{\bf t}, {\bf r}, \theta, \phi} ~ & ~ \left| h({\bf t}, {\bf r}, \theta, \phi)\right|^2  \\
		{\rm s.t.} ~ & ~ {\bf t} \in {\cal C}_{\rm t},~{\bf r} \in {\cal C}_{\rm r}, \\
		& ~ \theta,\phi \in [0,2\pi].
	\end{align}
\end{subequations}
Problem \eqref{opt_problem} is quite challenging to solve because the objective function is non-concave and has a complicated expression due to the exponential terms in the FRVs and PFVs shown in \eqref{sum_h}.  Besides, the coupling between the optimization variables adds to the complexity of the problem.  To overcome these challenges, we will develop an alternating optimization approach, which iteratively optimizes one of the variables while keeping the others fixed, thereby simplifying the problem into more tractable subproblems.

\section{Proposed Solution} \label{sec3}
In this section, we develop an alternating optimization approach that iteratively optimizes the transmit variables $\{{\bf r},\phi\}$ and the receive variables $\{{\bf t},\theta\}$ to maximize the channel power gain in problem \eqref{opt_problem}.

First, we optimize $\{{\bf r},\phi\}$ with given $\{{\bf t},\theta\}$.  In this subproblem, problem \eqref{opt_problem} can be rewritten as
\begin{subequations}
	\label{opt_subproblem}
	\begin{align}
		\max_{{\bf r},\phi} ~ & ~ |h({\bf r},\phi)|^2\\
		{\rm s.t.} ~ & ~ {\bf r} \in {\cal C}_{\rm r},~\phi\in[0,2\pi].
	\end{align}
\end{subequations}
By denoting $g({\bf r}, \phi) = |h({\bf r}, \phi)|^2$, the objective function in problem \eqref{opt_subproblem} can be further derived as
\begin{align} \label{g}
g({\bf r}, \phi) &=  {\bf g}({\bf r}, \phi)^H {\bf B} {\bf g}({\bf r}, \phi) \nonumber\\
& = \sum_{p = 1}^{2L_{\rm r}} [{\bf B}]_{pp} + 2\sum_{p<q} |[{\bf B}]_{pq}| \cos (\varpi_{pq}({\bf r}, \phi)),
\end{align}
where ${\bf B} = {\bf \Lambda} {\bf f}({\bf t}, \theta) {\bf f}({\bf t}, \theta)^H {\bf \Lambda}^H \in \mathbb{C}^{2L_{\rm r}\times 2L_{\rm r}}$ is a Hermitian matrix, and $\varpi_{pq}({\bf r}, \phi) = \omega_q({\bf r}, \phi) - \omega_p({\bf r}, \phi) + \angle [{\bf B}]_{pq}$ with 
\begin{equation}
  \omega_l({\bf r},\phi) = \begin{cases}
  \frac{2\pi}{\lambda} \rho_{{\rm r}, \frac{l+1}{2}}({\bf r}), \hspace{5.6mm} \text{$l$ is odd} \\
  \frac{2\pi}{\lambda} \rho_{{\rm r}, \frac{l}{2}}({\bf r}) + \phi, \hspace{2mm} \text{otherwise}
  \end{cases}\hspace{-3mm},~1 \le l \le 2L_{\rm r}.
\end{equation}
From \eqref{g}, it can be shown that the objective function $g({\bf r},\phi)$ is non-concave, as it involves cosine functions that depend nonlinearly on the receive APV ${\bf r}$ and phase shift $\phi$.

To address this challenge, we construct a quadratic surrogate function using a second-order Taylor expansion to locally approximate the objective function and adopt the SCA technique to iteratively optimize the receive variables $\{{\bf r},\phi\}$.  The gradient vector $\nabla g ({\bf r}, \phi) \in \mathbb{R}^3$ and Hessian matrix $\nabla^2 g ({\bf r}, \phi) \in \mathbb{R}^{3\times 3}$ of $g({\bf r}, \phi)$ over $\{{\bf r},\phi\}$ are necessary to this approximation, derived as follows.  First, $\nabla g ({\bf r}, \phi) = \left[ \frac{\partial g({\bf r}, \phi)}{\partial x_{\rm r}}, \frac{\partial g({\bf r}, \phi)}{\partial y_{\rm r}}, \frac{\partial g({\bf r}, \phi)}{\partial \phi} \right]^T$ can be derived as
\begin{align}
	\frac{\partial g({\bf r}, \phi)}{\partial x_{\rm r}} &= - \sum_{p < q} c_{pq} \left|[{\bf B}]_{pq}\right|  \sin (\varpi_{pq}({\bf r}, \phi)), \\
	\frac{\partial g({\bf r}, \phi)}{\partial y_{\rm r}} &= - \sum_{p < q} d_{pq }\left|[{\bf B}]_{pq}\right|  \sin (\varpi_{pq}({\bf r}, \phi)), \\
	\frac{\partial g({\bf r}, \phi)}{\partial \phi} &= - \sum_{p < q} \Upsilon_{pq} \left|[{\bf B}]_{pq}\right| \sin (\varpi_{pq}({\bf r}, \phi)),
\end{align}
where $c_{pq} = \frac{2\pi}{\lambda} (\cos\alpha_{{\rm r}, q} \sin\beta_{{\rm r}, q} - \cos\alpha_{{\rm r}, p}\sin\beta_{{\rm r}, p})$, $d_{pq} = \frac{2\pi}{\lambda} (\sin\alpha_{{\rm r}, q} - \sin\alpha_{{\rm r}, p})$, and
\begin{align}
  \Upsilon_{pq} = &\begin{cases}
  1, \hspace{4.7mm} \text{$p$ is odd, $q$ is even} \\
  -1, \hspace{2mm} \text{$p$ is even, $q$ is odd} \\
  0, \hspace{4.7mm} \text{otherwise}
  \end{cases}\hspace{-3mm},\nonumber \\
  &\hspace{15mm}1 \le p \le 2L_{\rm r},~1 \le q \le 2L_{\rm r}.
\end{align}
To ensure the convexity of the surrogate function, we further derive a curvature bound $\delta$ that satisfies $\delta{\bf I}_3 \succeq \nabla^2 g ({\bf r}, \phi)$ to guarantee the quadratic approximation remains globally conservative.  Following \cite[Appendix B]{MZZ24Apr}, we establish an upper bound on the spectral norm of Hessian matrix through its Frobenius norm, which leads to $\delta = \sqrt{1 + \frac{64\pi^4}{\lambda^4} + \frac{16\pi^2}{\lambda^2}} \sum_{p<q} \left| [{\bf B}]_{pq}\right|$.  Denoting $\tilde{\bf r} = [{\bf r}^T, \phi]^T$, we construct a quadratic surrogate function $\bar{g}(\tilde{\bf r})$ that satisfies
\begin{align} \label{surrg}
	g({\bf r}, \phi) 	&\ge \bar{g}(\tilde{\bf r}) = g(\tilde{\bf r}^{(i)}) + \nabla g(\tilde{\bf r}^{(i)})^T \left( \tilde{\bf r} - \tilde{\bf r}^{(i)} \right) \nonumber\\
	&\hspace{25mm} - \frac{\delta}{2} \left( \tilde{\bf r} - \tilde{\bf r}^{(i)} \right)^T\left( \tilde{\bf r} - \tilde{\bf r}^{(i)} \right) \nonumber\\
	&= \underbrace{-\frac{\delta}{2} \tilde{\bf r}^T \tilde{\bf r} + \left( \nabla g(\tilde{\bf r}^{(i)}) + \delta \tilde{\bf r}^{(i)} \right)^T \tilde{\bf r}}_{\tilde{g}(\tilde{\bf r})}\nonumber\\
	&\hspace{25mm}+ \underbrace{g(\tilde{\bf r}^{(i)}) - \frac{\delta}{2}(\tilde{\bf r}^{(i)})^T \tilde{\bf r}^{(i)}}_{\text{constant}}.
\end{align}
The surrogate function in \eqref{surrg} provides a global lower bound for the original objective function based on the candidate solution $\tilde{\bf r}^{(i)}$ at the $i$-the SCA iteration.

\begin{algorithm}[!t]
	\caption{SCA-based Solution for Solving Problem \eqref{opt_problem}}
	\label{alg}
	\footnotesize
	\renewcommand{\algorithmicrequire}{\textbf{Input:}}
	\renewcommand{\algorithmicensure}{\textbf{Output:}}
	\begin{algorithmic}[1]
		\REQUIRE $P_{\rm t}$, $\sigma$, ${\bf \Lambda}$, $\epsilon_1$, $\epsilon_2$, $L_{\rm t}$, $L_{\rm r}$, $I_{\max}$, $T_{\max}$, $\{\alpha_{{\rm t}, \iota}\}_{\iota=1}^{L_{\rm t}}$, $\{\beta_{{\rm t}, \iota}\}_{\iota=1}^{L_{\rm t}}$, $\{\alpha_{{\rm r}, \ell}\}_{\ell=1}^{L_{\rm r}}$, $\{\beta_{{\rm r}, \ell}\}_{\ell=1}^{L_{\rm r}}$, ${\cal C}_{\rm t}$, ${\cal C}_{\rm r}$, $\lambda$, ${\bf t}_0$, ${\bf r}_0$.
		\ENSURE ${\bf t}$, ${\bf r}$, $\theta$, $\phi$.
		\STATE Initialize ${\bf t}^{(0)} = {\bf t}_0$, ${\bf r}^{(0)} = {\bf r}_0$, $\theta^{(0)} = 0$, and $\phi^{(0)} = 0$.
		\FOR{$t = 1 \rightarrow T_{\max}$}
		\STATE Update ${\bf t}^{(t)} \leftarrow {\bf t}^{(t-1)}$, ${\bf r}^{(t)} \leftarrow {\bf r}^{(t-1)}$, $\theta^{(t)} \leftarrow \theta^{(t-1)}$, $\phi^{(t)} \leftarrow \phi^{(t-1)}$.
		\STATE Initialize $\tilde{\bf r}^{(0)}$ using ${\bf r}^{(t)}$ and $\phi^{(t)}$ for SCA.
		\FOR{$i = 1 \rightarrow I_{\max}$}
		\STATE Update $\tilde{\bf r}^{(i)} \leftarrow \tilde{\bf r}^{(i-1)}$.
		\STATE Compute ${\bf B}$ and $\nabla g(\tilde{\bf r}^{(i)})$.
		\STATE Obtain $\tilde{\bf r}^{(i+1)}$ by solving \eqref{quad_problem1}.
		\IF{Increase of the objective function in \eqref{quad_problem1} is below $\epsilon_2$}
		\STATE Break.
		\ENDIF
		\ENDFOR
		\STATE Update ${\bf r}^{(t)}$ and $\phi^{(t)}$ via $\tilde{\bf r}^{(i+1)}$.
		\STATE Initialize $\tilde{\bf t}^{(0)}$ using ${\bf t}^{(t)}$ and $\phi^{(t)}$ for SCA.
		\FOR{$i = 1 \rightarrow I_{\max}$}
		\STATE Update $\tilde{\bf t}^{(i)} \leftarrow \tilde{\bf t}^{(i-1)}$.
		\STATE Compute ${\bf D}$ and $\nabla f(\tilde{\bf t}^{(i)})$.
		\STATE Obtain $\tilde{\bf t}^{(i+1)}$ by solving \eqref{quad_problem2}.
		\IF{Increase of the objective function in \eqref{quad_problem2} is below $\epsilon_2$}
		\STATE Break.
		\ENDIF
		\ENDFOR
		\STATE Update ${\bf t}^{(t)}$ and $\theta^{(t)}$ via $\tilde{\bf t}^{(i+1)}$.
		\IF{Increase of the achievable rate in \eqref{rate} is below $\epsilon_1$}
		\STATE Break.
		\ENDIF
		\ENDFOR
		\STATE Set ${\bf t} = {\bf t}^{(t)}$, ${\bf r} = {\bf r}^{(t)}$, $\theta = \theta^{(t)}$, and $\phi = \phi^{(t)}$.
		\RETURN ${\bf t}$, ${\bf r}$, $\theta$, $\phi$.
	\end{algorithmic}
\end{algorithm}

Therefore, the original maximization of $g(\mathbf{r}, \phi)$ can be reformulated as maximizing its concave quadratic approximation $\tilde{g}(\tilde{\mathbf{r}})$ from \eqref{surrg}.  This yields the following convex problem at the $i$-th SCA iteration, i.e.,
\begin{subequations}
	\label{quad_problem1}
	\begin{align}
		\max_{\tilde{\bf r}} ~ & - \frac{\delta}{2} \tilde{\bf r}^T \tilde{\bf r} + \left( \nabla g(\tilde{\bf r}^{(i)}) + \delta \tilde{\bf r}^{(i)} \right)^T \tilde{\bf r} \\
		{\rm s.t.} ~ & ~ {\bf r} \in {\cal C}_{\rm r},~\phi\in[0,2\pi]. \label{cons_quad}
	\end{align}
\end{subequations}
Problem \eqref{quad_problem1} is a quadratic programming (QP) problem and can be efficiently solved by standard tools like quadprog \cite{TW20Jul}.

Next, we optimize $\{{\bf t}, \theta\}$ with given $\{{\bf r}, \phi\}$.  In this subproblem, we denote ${\bf D} = {\bf \Lambda}^H {\bf g}({\bf r}, \phi) {\bf g}({\bf r}, \phi)^H {\bf \Lambda} \in \mathbb{C}^{2L_{\rm t}\times 2L_{\rm t}}$, $\tilde{\bf t} = [{\bf t}^T, \theta]^T$, and $f(\tilde{\bf t}) = {\bf f}({\bf t},\theta)^H {\bf D} {\bf f}({\bf t},\theta)$.  Similar to the procedure for optimizing $\{{\bf r}, \phi\}$, problem \eqref{opt_problem} can be converted to
\begin{subequations}
	\label{quad_problem2}
	\begin{align}
		\max_{\tilde{\bf t}} ~ & - \frac{\delta}{2} \tilde{\bf t}^T \tilde{\bf t} + \left( \nabla f(\tilde{\bf t}^{(i)}) + \delta \tilde{\bf t}^{(i)} \right)^T \tilde{\bf t} \\
		{\rm s.t.} ~ & ~ {\bf t} \in {\cal C}_{\rm t},~\theta\in[0,2\pi], \label{cons_quad}
	\end{align}
\end{subequations}
by replacing $\left\{ L_{\rm r}, {\cal C}_{\rm r}, {\bf r}, \phi, {\bf B}, \{\alpha_{{\rm r}, \ell}\}_{\ell=1}^{L_{\rm r}}, \{\beta_{{\rm r}, \ell}\}_{\ell=1}^{L_{\rm r}} \right\}$ with $\left\{ L_{\rm t}, {\cal C}_{\rm t}, {\bf t}, \theta, {\bf D}, \{\alpha_{{\rm t}, \iota}\}_{\iota=1}^{L_{\rm t}}, \{\beta_{{\rm t}, \iota}\}_{\iota=1}^{L_{\rm t}} \right\}$, which is also a convex optimization problem that can be efficiently solved.

Algorithm \ref{alg} is presented to iteratively solve problem \eqref{opt_problem}, which begins by initializing the APVs at ${\bf t}^{(0)} = {\bf t}_0$ and ${\bf r}^{(0)} = {\bf r}_0$, with initial phase shifts $\theta^{(0)} = 0$ and $\phi^{(0)} = 0$.  To ensure robustness, we test $\zeta$ uniformly distributed starting positions across the moving region and select the position with the highest achievable rate.  The algorithm continues iterating until either the maximum iteration count $T_{\max}$ is reached or the rate increment falls below $\epsilon_1$.  Each iteration performs SCA for the transmit variables $\{{\bf t},\theta\}$ and receive variables $\{{\bf r},\phi\}$, with each SCA loop running for up to $I_{\max}$ iterations or until the increment of the objective function is below $\epsilon_2$.  The final values of ${\bf t}^{(t)}$, ${\bf r}^{(t)}$, $\theta^{(t)}$ and $\phi^{(t)}$ are returned as the solution.

Furthermore, the convergence of Algorithm \ref{alg} is guaranteed since the alternating optimization process ensures that the achievable rate never decreases between iterations and the achievable rate has a finite maximum limit.  We will also demonstrate this convergence behavior through simulations in Section \ref{sec4}.  Finally, the total computational complexity of the algorithm is given by ${\cal O}\left( \zeta T_{\max} I_{\max} \left( L_{\rm r}^2 L_{\rm t} + L_{\rm t}^2 L_{\rm r} + L_{\rm r}^2 + L_{\rm t}^2 \right) \right)$, which is polynomial over $L_{\rm t}$ and $L_{\rm r}$.

\section{Simulation Results} \label{sec4}
\begin{figure}[!t]
	\centering
	\includegraphics[width=0.66\linewidth]{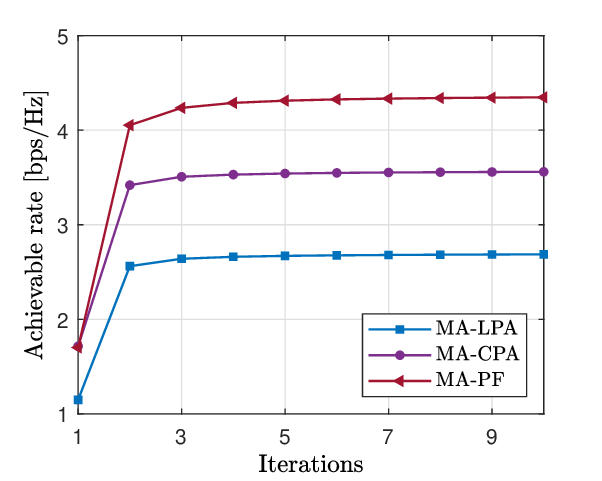}
	\caption{Convergence behavior of Algorithm \ref{alg}.}
	\label{fig_conv}
\end{figure}

We present extensive simulation results in this section to evaluate the performance gains of the proposed system over conventional systems and validate the effectiveness of the proposed algorithm.

In the simulations, we assume that the PSs for polarforming can be continuously tuned from $0$ to $2\pi$.  The polarized channel matrix is modeled as ${\bf \Lambda}_{\ell \iota} = {\bf \Psi} \odot {\bf H}^{\rm i.i.d.}_{\ell\iota}$ \cite{ZDW24Sep} with
\begin{equation}
  {\bf \Psi} = \frac{1}{\sqrt{\chi + 1}} \begin{bmatrix} 1 & \sqrt{\chi} \\ \sqrt{\chi} & 1 \end{bmatrix},
\end{equation}
where $\chi$ is the inverse cross-polarization discrimination (XPD) to quantify the degree of channel depolarization, and ${\bf H}^{\rm i.i.d.}_{\ell\iota}$ consists of independent circularly symmetric complex Gaussian elements with a normalized variance of one.  Unless otherwise stated, the inverse XPD is set to $\chi = 1$.  We consider geometric channels with $L_{\rm t} = L_{\rm r} = L$, where the PPRM in \eqref{pprm} follows ${\bf \Lambda} = \frac{1}{\sqrt{\kappa+1}} {\rm blkdiag} \left\{\sqrt{\kappa}{\bf \Lambda}_{11}, \frac{1}{\sqrt{L-1}}{\bf \Lambda}_{22}, \cdots, \frac{1}{\sqrt{L-1}}{\bf \Lambda}_{LL}\right\}$, with $\kappa$ being the Rician factor.  We aslo assume $A_{\rm t} = A_{\rm r} = A$.    Due to channel normalization, the average signal-to-noise ratio (SNR) of the proposed system is only determined by the transmit power $P_{\rm t}$ and the noise power $\sigma^2$ at the receiver, i.e., $\frac{P_{\rm t}}{\sigma^2}$.  For Algorithm~\ref{alg}, we set the convergence thresholds to $\epsilon_1 = \epsilon_2 = 10^{-6}$, with maximum iteration limits of $T_{\max} = 20$ and $I_{\max} = 800$.  The multiple initialization process uses $\zeta = 4 \left\lceil \frac{A}{\lambda} \right\rceil^2$ to ensure robust convergence, where $\lceil \cdot \rceil$ denotes the ceiling function that rounds up to the nearest integer.  Besides, all results are obtained by averaging over $10^4$ independent Monte Carlo channel realizations.

In comparison to the proposed scheme, we consider benchmark schemes from two different aspects: 1) antenna position configurations, by employing FPAs or MAs, marked by ``FPA'' and ``MA'', respectively; and 2) antenna polarization configurations \cite{ZDW24Sep}, by applying linearly polarized antennas (LPAs), circularly polarized antennas (CPAs), DPAs, or the polarforming technique, marked by ``LPA'', ``CPA'', ``DPA'', and ``PF'', respectively.  In particular, the proposed scheme, marked by ``MA-PF'', enables both the transmit and receive antennas to flexibly change positions and adopt polarforming.  Other benchmark schemes are defined similarly with other possible antenna position/polarization configuration pairing.

First, Fig.\;\ref{fig_conv} illustrates the convergence behavior of Algorithm \ref{alg} for different polarization configurations by setting $\text{SNR} = 5$ dB, $L = 6$, $A = \lambda$, and $\kappa = 0$ dB.  It is observed that the achievable rate consistently increases and reaches a maximum value after six iterations for all polarization configurations, which validates the convergence analysis in Section \ref{sec3}.

\begin{figure}[!t]
	\centering
	\subfigure[]{
		\centering \label{fig_compare_DPA}
		\includegraphics[width=0.48\columnwidth]{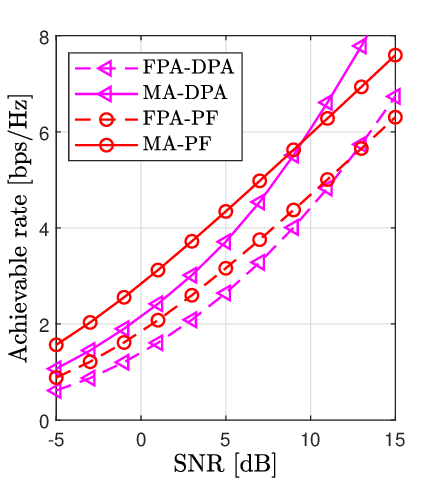}}
	\subfigure[]{
		\centering \label{fig_compare_FPA}
		\includegraphics[width=0.48\columnwidth]{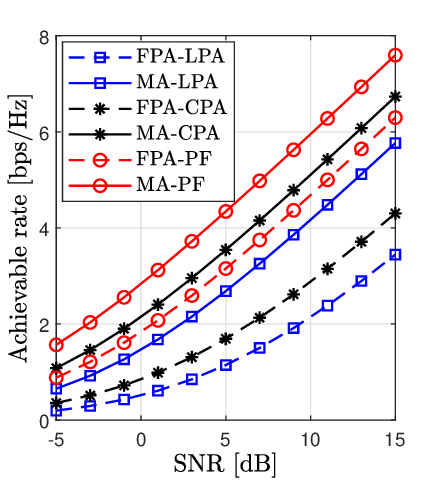}}
	\caption{Achievable rate versus SNR.}
	\label{fig_rate_snr}
\end{figure}

Next, Fig.\;\ref{fig_rate_snr} compares the achievable rate of the proposed scheme with benchmark schemes, where the parameters are set to $L = 6$, $A = \lambda$, and $\kappa = 0$ dB.  In Fig.\;\ref{fig_compare_DPA}, the achievable rate of the MA-DPA scheme is found by the exhaustive search, where the moving region is divided into a $20 \times 20$ grid, and the maximum rate for each grid point is achieved using water-filling as in \cite{ZDW24Sep}.   At low SNR, the DPA scheme achieves a lower rate than the PF scheme; however, it outperforms the PF scheme at higher SNR due to multiplexing gains from two RF chains and two data streams.  From Fig.\;\ref{fig_compare_FPA}, the MA-based schemes achieve better performance than the FPA-based schemes, and those by polarforming outperform the schemes with fixed polarizations.  Remarkably, the proposed scheme achieves the best performance among all benchmark schemes.  This implies that antenna positions and polarizations can be individually optimized to enhance system performance, as they leverage distinct spatial and polarization DoFs.

\begin{figure*}[!t]
	\centering
	\subfigure[$A = \lambda$ and $\kappa = 0$ dB.]{
		\centering \label{fig_paths}
		\includegraphics[width=0.66\columnwidth]{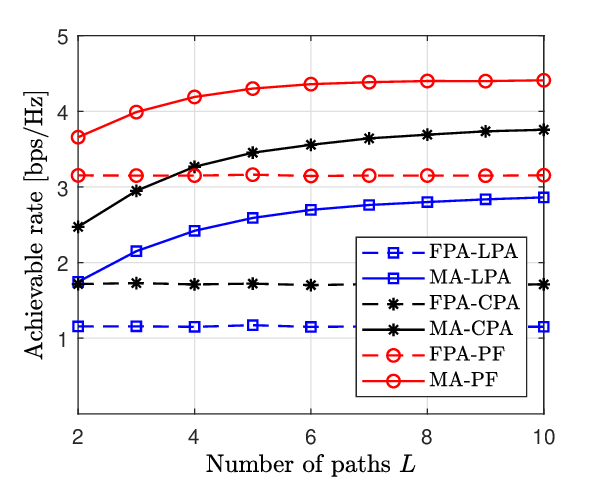}}
	\subfigure[$L = 6$ and $\kappa = 0$ dB.]{
		\centering \label{fig_region}
		\includegraphics[width=0.66\columnwidth]{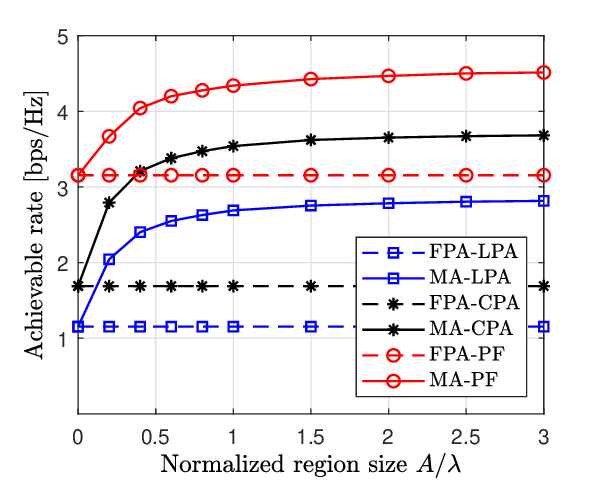}}
		\subfigure[$L = 6$ and $A = \lambda$.]{
		\centering \label{fig_kappa}
		\includegraphics[width=0.66\columnwidth]{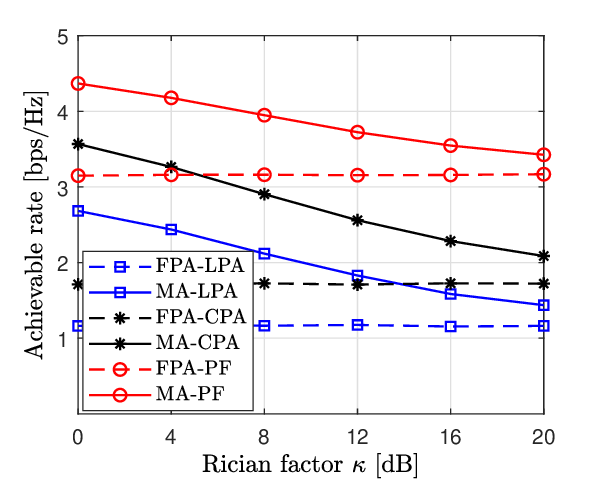}}
	\caption{Impact of the number of paths, moving region size, and Rician factor on the performance of the considered schemes.}
	\label{fig_impact}
\end{figure*}

Finally, Fig.\;\ref{fig_impact} investigates the impacts of the number of paths, moving region size, and Rician factor on the performance of different schemes at a fixed SNR of $5$ dB.  From Figs. \ref{fig_paths} and \ref{fig_region}, the achievable rates of MA-based schemes exhibit a consistent improvement as either the number of paths or normalized moving region size increases.  This trend holds across all polarization configurations and underscores the inherent advantage of MAs in exploiting spatial DoFs.  Moreover, polarforming delivers substantial performance gains over conventional fixed-polarization schemes, regardless of whether the antenna employs a fixed or flexible position.  This improvement arises from the capability of polarforming in dynamically aligning the antenna polarization with that of EM waves, thereby maximizing the received signal power while mitigating polarization loss.  In Fig.\;\ref{fig_kappa}, the performance gap between MA-based and FPA-based schemes diminishes as the Rician factor increases, as the benefits of spatial DoFs via antenna movement become less pronounced in the presence of a dominant path.  However, even in this case, polarforming still provides significant performance enhancements by optimizing the polarization alignment with the dominant path, thus demonstrating its robustness across diverse channel conditions.

\section{Conclusion} \label{sec5}
In this letter, we proposed and investigated polarforming design for the MA-enabled communication system that enables flexible antenna positions and polarizations to fully exploit both spatial and polarization DoFs.  We then developed an SCA-based optimization algorithm to maximize the achievable rate of the proposed system by iteratively tuning the antenna positions and phase shifts at both the transmitter and receiver.  Furthermore, numerical results demonstrated the superior performance of the proposed system over conventional systems deploying fixed positions and/or fixed polarizations by more effectively mitigating channel depolarization and adapting to channel fading.

\end{document}